# The stability of graphene based Möbius strip with vacancy at high temperature

Kaishuai Yang [a,b], Chuanguo Zhang [a,b], Xiaohong Zheng [a,b], Xianlong Wang [a,b,*], Zhi Zeng [a,b,*]

[a] *Key Laboratory of Materials Physics, Institute of Solid State Physics, Chinese Academy of Sciences, Hefei 230031, China*

[b] *University of Science and Technology of China, Hefei 230026, China*

*Corresponding authors

Xianlong Wang

E-mail: xlwang@theory.issp.ac.cn

TEL: +86-551-65591150

Zhi Zeng

E-mail: zzeng@theory.issp.ac.cn

TEL: +86-551-65591407

# ABSTRACT

By using the density functional theory (DFT) combined with the molecular dynamics (MD) simulations, structural and electronic properties of mono-vacancy (MV) defect in Möbius strip formed from graphene are investigated. Two kinds of MV are observed depending on the local structures around defects. At static condition, in the curved areas of Möbius strip, MV has the configuration of one pentagon and one nonagon ring (59-type), which is similar to that of carbon nanotubes and graphene. the most stable MV appear in the twisted areas and has a profile of two pentagon and two hexagon rings (5566-type) with one $sp^3$ hybridized carbon at the central site. While DFT-MD simulations prove that the 5566-type MV is an unstable configuration at room temperature and will transform into a 59-type MV. Additionally, the melting behavior of graphene based Möbius strips are investigated through empirical potential MD simulations, and we find that their melting temperature is about 2750 K, which is lower than that of carbon nanotubes and graphene.

# 1. Introduction

The graphene based Möbius strip (GMS) has attracted extensive attentions due to its special topological property, namely, with only one face and one edge, which makes it distinct from other carbon based low-dimensional materials, such as carbon nanotubes and graphene. Möbius aromatic hydrocarbon had been successfully synthesized in experiments,[1, 2] while the GMS has not been experimentally synthesized yet. Moreover, theoretical simulations focused on the properties of GMS heavily, since it may have important applications.[3-11] The simulated results demonstrate that GMS is a stable structure,[9] and show novel optical[4, 10] and magnetic properties.[6, 10] Especially, it is also believed to be a topological insulator.[7] In our previous work,[9] the structural configurations of GMS with different ratios between length and width are presented, and we find that since GMS has only one edge, differing from graphene nanoribbons, ferromagnetic state is the ground sate of GMS.

Carbon based materials are usually not ideal crystals, and atomic vacancies can affect their properties significantly,[12-21] for example, inducing magnetism in graphene.[13, 15] Previous works focused heavily on the properties of atomic vacancies presented in carbon nanotubes and graphene,[12-15, 17-30] where atomic vacancies can be introduced unintentionally during the processes of synthesis or deliberately by irradiation, chemical, and plasma treatments. Among different types of atomic vacancies in carbon nanostructures, mono-vacancy (MV) with one atom missing from lattice is a simple, popular and attractive one, and it has been identified clearly by experimentalists and theorists in carbon nanotubes and graphene.[15, 29-32] In these systems, MV undergoes a Jahn-Teller distortion, which leads to the formation of covalent bond between two of three atoms located around the atomic vacancy resulting in one five-membered ring and one nine-membered ring (59-type MV). To date, only 59-type MV was reported in carbon nanotubes and graphene.

To fully understand the properties of GMS, it is important and necessary to illustrate the MV properties of GMS. First of all, the structural configurations of the MV in the GMS should be clarified. As shown in our previous work,[9] since each GMS is composed by two kinds of areas, curved and chiral twisted parts, the MV features in GMS should be investigated separately depending on its local structures. In this work, based on the first-principles method, the structural and electronic properties of the MV in GMS are investigated. Our results show that besides 59-type MV present in the curved area, a 5566-type MV, which has two pentagons and two hexagons and acquires one $sp^3$ carbon atom at the central site, is observed in the twisted region. To the best of our knowledge, the 5566-type MV are not reported in carbon based low-dimensional materials. Furthermore, due to the strain introduced by their edges, graphene nanoribbons with chiral twist were theoretically predicted,[33-37] and in fact experimentally observed.[38] the shapes of graphene nanoribbons with different chiral twisted degrees were clearly shown by transmission electron miscroscopy (TEM) images[39]. Furthermore, graphene nanoribbons can transform into carbon nanotubes through chiral twisting,[40] and the local configurations of the twisted areas of GMS are similar to that of graphene nanoribbons with chiral twisting. Therefore, the explored MV behaviors of present work can also be used to analyze the properties of chirally twisted carbon nanotubes and graphene. Furthermore, the melting temperature of GMSs with different ratios between length and width is illustrated.

## 2. Computational methods

The calculations are performed by the SIESTA package, in which the normconserving pseudopotential and linear combinations of atomic orbital basis sets are used.[41, 42] The wave function is expanded with a double-ζ (DZ) basis sets, and the exchange-correlation potential of generalized gradient approximation (GGA) with the form of PerdewBurke-Ernzerhof (PBE) is selected.[43] The lattice vectors (50×50×50 Å) are large enough to avoid the interactions from adjacent neighbors. All related structures are fully relaxed. The method and parameters used in this work are basically the same as that in our previous works.[9, 44] The melting behaviors of GMSs are investigated via MD simulations using the LAMMPS[45] and Adaptive Intermolecular Reactive Empirical Bond Order (AIREBO) Potential[46] for a system of carbon and hydrogen atoms with the LJ cutoff 3.0 Å. The configurations of GMSs are relaxed under a series of Nose-Hoover thermostat[46, 47] from 1000 K to 3500 K for $6.25 \times 10^6$ steps with a time step of 0.5 fs.

## 3. Results and discussion

In our previous work,[9] the structural features and formation energies of GMS as a function of width-to-length ratio are characterized. We find that one, two, and three planar triangle regions can be observed in GMS, and their formation energies increase with the ratio of width-to-length increasing. Following, we will use *n-m* to represent the GMS structures, where *n* and *m* indicate the length (numbers of armchair lines) and the width (numbers of zigzag lines) of GMS, respectively. For example, 30-06 means the GMS with the length of 30 armchair lines and the width of 6 zigzag lines, which contains one planar triangle area, as shown in Fig. 1(a). Because of its topological property, GMS can be separated to three different areas depending on local deformation. As shown in Fig. 1(a), A and C denote the areas with twisting, while B is the curved area similar to carbon nanotubes. One carbon atom in the center site of A, B, and C areas is removed to create the MV, and these MVs are hereafter denoted as MV-A, MV-B, and MV-C, respectively. The formation energy ($\Omega$) of MV in GMS (labeled as $\Omega_{MV}$(GMS)) and graphene (labeled as $\Omega_{MV}$(G)) are given by the following equations:

$$\Omega_{MV}(GMS) = E_{MV}(GMS) + E_{atom} - E(GMS) \quad (1)$$

$$\Omega_{MV}(G) = E_{MV}(G) + E_{atom} - E(G) \quad (2)$$

$$E_{atom} = E(G)/N \quad (3)$$

where $E$(GMS) ($E$(G)) and $E_{MV}$(GMS) ($E_{MV}$(G)) denote the total energies of GMS (graphene) without and with MV, respectively, and $N$ is the number of carbon atoms in a 12×12 graphene supercell. The calculated results are presented in Table 1, the formation energy of MV in graphene is 7.56 eV, which agrees well with that of the published results (7.4 eV[32] and 7.7 eV[15]), indicating that our used method and parameters can sufficiently describe graphene. The formation energies of MV-A, MV-B, and MV-C are 6.43 eV, 6.51 eV and 6.72 eV, respectively. The MV formation energies of GMS are about 1 eV smaller than that of graphene (7.56 eV). The results show that compared with infinite graphene, it will be easier to introduce MVs in GMS or graphene with chiral twisting partially due to the existence of strain. Furthermore, the most stable site for MV in GMS locates at the planar triangle area (A area), since the formation energy of MV-A is the smallest one, as shown in Table 1.

The relaxed MV structures are presented in Fig. 1(b-d), and the zoomed in views are shown in the corresponding insets to clearly illustrate their structures. As can be found in Fig. 1(c), similar to the MV of carbon nanotubes and graphene,[15, 31, 32], MV-B shows a typical 59-type pattern with one pentagon and one enneagon. A new bond is formed between atom II and atom III with the bond-length of 1.59 Å, which is much shorter by 0.98 Å than the distance (2.57 Å) between atom I and atom II, due to the Jahn-Teller distortion. Interestingly, MV-A and MV-C in the twisted areas finally evolve into a novel 5566-type pattern, where the central carbon atom has $sp^3$-type hybridization after full relaxation. The structural properties of MV-A and MV-C are shown in Table 2, and the results show that the averaged distance between the central carbon and its four neighbors is about 1.64 Å, which is significantly longer than the C-C bond-length (1.42 Å) of graphene and the new bond formed in MV-B (1.59 Å) and diamond (1.55 Å). Combined with the bond-angle information of MV-A and MV-C shown in Table 2, we can conclude that the carbon atoms located at their central sites are of a $sp^3$ type.

To further confirm this, the electronic structures of MVs are studied. Fig. 2(a) and (b) present the partial density of states (PDOS) of atom at the edge and at the central site of the ideal planar triangle area in GMS, respectively. Meanwhile, the PDOS of the central carbon atom in 5566-type MV is presented in Fig. 2(c). The carbon atom at the edge sites of GMS is spin polarized, which is similar to the case of graphene nanoribbons with zigzag edge and agrees with previous work.[9, 44] Furthermore, if we compare Fig. 2(b) with 2(c), the energy gap around the Fermi level shown in Fig. 2(c) is much larger than that of $sp^2$ hybridized carbon in Fig. 2(b), indicating that the central carbon atom of MV-A and MV-C hybridize with its four neighbors by $sp^3$-like bonding. In other words, at low temperature condition, the 5566-type MV can contribute $sp^3$-type spectral signals in the carbon based low-dimensional materials with chiral twisting.

Following, we will investigate the stability of 5566-type MV-A at room temperature, which is the most stable MV configuration in the static condition. For this purpose, DFT-MD simulations of MV-A at room temperature (T = 300 K) is carried out in the time step of 1 fs, and the corresponding result is shown in Fig. 3(a). After running 8000 DFT-MD steps, we find that the 5566-type MV-A geometry (shown in Fig. 3(b)) finally transform into a 59-type MV-B pattern (shown in Fig. 3(c)). The result show that the 5566-type defect can only exist at low temperature and is unstable at room temperature. The 5566-type defect will undergo an immigration from the twisting area A to the curved area B, meanwhile turning into a 59-type defect. Similar scenario may also occur in other low-dimensional materials, such as carbon nanotubes and graphene.

Even though the melting temperature of graphene (~4500 K)[48] and carbon nanotubes (~3100 K)[49] are known, the melting temperature of GMS is still not clarified yet. The melting temperature of GMS is on of important parameters for its potential applications. To investigate the melting temperature of GMS, as shown in Fig. 4, five different GMS configurations (30-06, 30-08, 30-10, 30-12, 30-14) with different widths are investigated. The results of empirical potential MD simulations show that the internal energies of these GMSs have a sharp increase when the temperature increase to a certain value. The dramatic changes of internal energy are clear signal for the larger structural changes. We define the temperature, where the internal energies increase dramatically, as the melting point. Figure 4 illuminates that the melting points of these GMSs are almost the same (~2750 K), due to that the melting point is mainly determined by the materials' bonding property and internal stress. The results show the melting temperature of GMS is much lower than that of carbon nanotubles and graphene.

## 4. Conclusions

The structural and electronic properties of MV in Möbius strip formed from graphene are simulated based on the DFT. A MV defect with 5566-type pattern with a $sp^3$ hybridized carbon atom at central site is observed in the chiral twisted areas of GMS, while a 59-type MV appears in the curved area. Though the 5566-type MV has the lowest energy at 0 K, our DFT-MD simulations prove that the 5566-type MV is not stable at room temperature, and it will transform into a 59-type MV, the similar scenario may occur in other low-dimensional materials, e.g., carbon nanotubes. Additionally, the melting temperature of GMSs are about 2750 K, which is lower than that of carbon nanotubles and graphene.

## Acknowledgments

This work was supported by the National Science Foundation of China under Grant Nos. 11674329, 11605231, and 11574318, Science Challenge Project, No. TZ2016001. The calculations were partly performed in Center for Computational Science of CASHIPS.

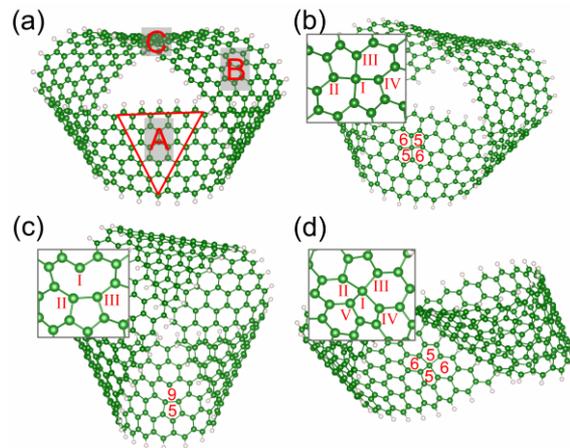

Fig. 1. (a) The structure of ideal Möbius strip, and A, B, and C are used to present the areas to create mono-vacancy (MV). Red triangle shows the planar area. (b), (c), and (d) show the relaxed structures of MV introduced in the A (MV-A), B (MV-B), and C (MV-C) areas, respectively, while the enlarged structures are presented in the insets. The Arabic numbers presented in figures are used to indicate the configurations of MV. The Greek numbers presented in insets are used to indicate the atomic sites of MV.

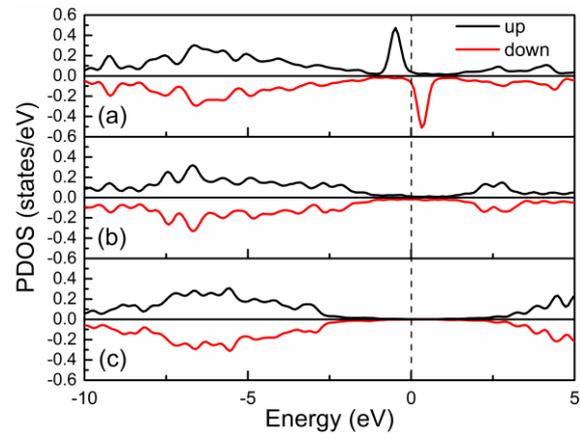

Fig. 2. The partial density of states (PDOS) of carbon atom located at the edge site of perfect Möbius strip (a), the central site of planar triangle area of perfect Möbius strip (b), and the central site of MV-A (c).

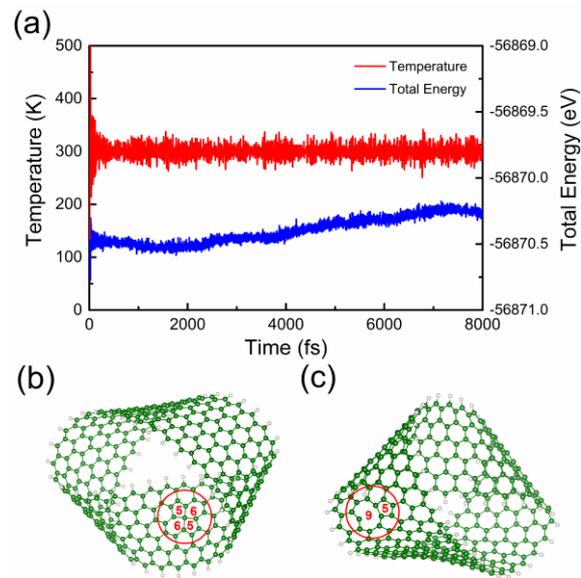

Fig. 3. (a) Temperature and total energy as a function of time obtained from molecular dynamics simulation. (b) and (c) present the initio (5566-type MV) and final (59-type MV) configuration before and after running the molecular dynamics simulation

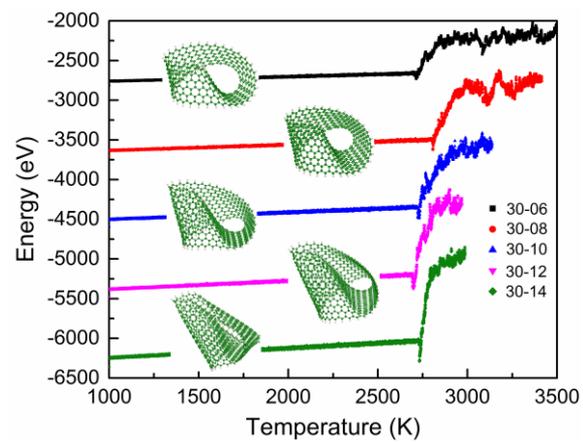

Fig. 4. Internal energies of Möbius strips numbered by 30-06, 30-08, 30-10, 30-12, and 30-14 are shown in the black, red, blue, pink, and green lines, respectively. The structures of the Möbius strips are shown in the corresponding insets.

Table 1

Formation energies of mono-vacancies.

| eV | graphene | MV-A | MV-B | MV-C |
|---|---|---|---|---|
| Formation energy | 7.56 | 6.43 | 6.51 | 6.72 |

Table 2

The structural properties of MV-A and MV-C. Distance (*d*) and angle (*a*) are shown in the unite of Å and degree, respectively.

|      | $d_{\text{I-II}}$ | $d_{\text{I-III}}$ | $d_{\text{I-IV}}$ | $d_{\text{I-V}}$ | $a_{\text{II-I-IV}}$ | $a_{\text{III-I-V}}$ | $a_{\text{II-I-III}}$ | $a_{\text{II-I-V}}$ |
|------|------|------|------|------|-------|-------|------|------|
| MV-A | 1.62 | 1.67 | 1.60 | 1.68 | 138.0 | 147.6 | 96.2 | 96.1 |
| MV-B | 1.61 | 1.69 | 1.60 | 1.68 | 138.1 | 145.7 | 96.3 | 96.1 |